\documentclass[prb,twocolumn,showpacs]{revtex4}
\usepackage{amsmath}
\usepackage{amssymb}
\usepackage{graphicx}
\usepackage{bm}
\usepackage{dcolumn}
\begin{document}

\title{Conductivity anisotropy helps to reveal the microscopic structure
of a density wave at imperfect nesting}
\author{P. D. Grigoriev}
\affiliation{L. D. Landau Institute for Theoretical Physics,
Chernogolovka, 142432, Russia} \affiliation{Institut
Laue-Langevin, Grenoble, France}
\author{S. S. Kostenko}
\affiliation{Institute of Problems of Chemical Physics, 142432,
Chernogolovka, Russia}

\begin{abstract}
Superconductivity or metallic state may coexist with density wave
ordering at imperfect nesting of the Fermi surface. In addition to
the macroscopic spatial phase separation, there are, at least, two
possible microscopic structures of such coexistence: (i) the
soliton-wall phase and (ii) the ungapped Fermi-surface pockets. We
show that the conductivity anisotropy allows to distinguish
between these two microscopic density-wave structures. The results
obtained may help to analyze the experimental observations in
layered organic metals (TMTSF)$_{2}$PF$_{6}$, (TMTSF)$_{2}$ClO$_{4}$, $%
\alpha $-(BEDT-TTF)$_{2}$KHg(SCN)$_{4}$ and in other compounds.
\end{abstract}

\maketitle

\section{Introduction}

At imperfect nesting, the charge/spin-density-wave gap in electron
spectrum does not cover the whole Fermi surface, so that the
density-wave and metallic/superconducting states may coexist. In
addition to the macroscopic spatial phase
separation,\cite{ChaikinPRL2014} such density-wave state may have
two different microscopic structures: (i) spatially uniform
structure with reconstructed Fermi surface (FS), containing
ungapped parts \cite{MoceauAdvPhys,GrigorievPRB2008} and (ii)
spatially non-uniform soliton structure
\cite{BrazKirovaReview,SuReview,GrigPhysicaB2009}.
Superconductivity, appearing on such density-wave background is
rather common \cite{GabovichReview,Vuletic} and has many unusual
properties, including the
strong enhancement of the upper critical field \cite%
{Hc2Pressure,CDWSC,GrigorievPRB2008,GrigPhysicaB2009}, anisotropic
transition temperature \cite{KangPRB2010} etc. The knowledge of the
microscopic structure of the density wave (DW) state at imperfect nesting is
important for describing various compounds, where charge/spin-density wave
coexists with metallic/superconducting states. For all above scenarios the
metallic conductivity decreases but does not vanish after the transition to
density-wave state. This decrease of conductivity may be anisotropic\cite%
{SinchGrigPRL2014} and depend on the DW microscopic structure. In the
present report we show how this anisotropy can help to distinguish various
DW microscopic structures. These results may be useful to describe the
electronic properties of organic metals $\alpha $-(ET)$_{2}$MHg(SCN)$_{4}$
with M = K and Tl, (TMTSF)$_{2}X$ with $X=$PF$_{6}$ or ClO$_{4}$, and of
other compounds.

For (TMTSF)$_{2}$PF$_{6}$ there are several arguments in favor of
a macroscopic phase separation \cite{ChaikinPRL2014} rather than
soliton \cite{BrazKirovaReview,SuReview,BGL,GL,GG,GGPRB2007}
scenario. The main of these
arguments is the constant (pressure independent) transition temperature $%
T_{c}^{SC}$ to the superconducting (SC) state, which is almost the
same in the uniform (high-pressure) and nonuniform (coexistent)
superconducting state. This is not the case for $\alpha
$-(ET)$_{2}$MHg(SCN)$_{4}$ organic metals, where the soliton-wall
scenario is more probable. The other arguments in favour of
macroscopic phase separation in (TMTSF)$_{2}$PF$_{6}$ in Ref.
\cite{ChaikinPRL2014} involve magnetic field, which changes the
soliton phase in the presence of imperfect nesting and may even
lead to a field-induced spin-density wave (FISDW) with soliton
microscopic structure. Although some theoretical results on the
evolution of the soliton DW structure in magnetic field have been
obtained for ideally 1D conductor \cite{BDK}, there is no
theoretical study of the soliton DW microscopic structure in
magnetic field in the presence of imperfect nesting. The NMR
absorption lineshape \cite{Brown2005} also suggests the soliton DW
structure rather than macroscopic domains with usual uniform DW.
Moreover, SC appearing in the soliton wall phase of spin-density
wave (SDW) has the triplet SC pairing \cite{GGPRB2007} in
agreement with experiments in
(TMTSF)$_{2}$PF$_{6}$.\cite{LeeTripletMany} From energy
considerations, the soliton wall scenario
\cite{BrazKirovaReview,SuReview,BGL,GL,GG,GGPRB2007} is more
favorable, because the energy loss $W$ due to nonuniform DW order
parameter is compensated by the large kinetic energy of
soliton-band quasiparticles \cite{BGL,GG}. Thus, the question
about the DW structure in (TMTSF)$_{2}$PF$_{6}$ in the coexistence
region is still open. The "double" spatial separation is also
possible, where there are macroscopic domains of metallic and
soliton DW states.

\section{Temperature dependence of conductivity anisotropy in quasi-1D
conductors with imperfect nesting}

\begin{figure}[tb]
\includegraphics[width=0.49\textwidth]{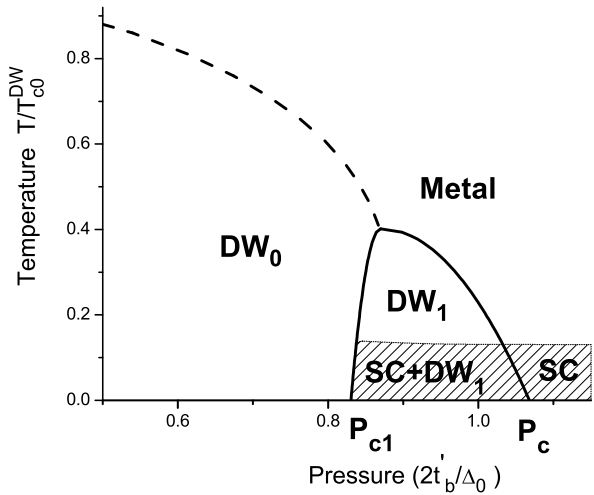}
\caption{The schematic picture of the phase diagram in organic metals, where
metal/superconducting state coexists with DW in some pressure interval above
$P_{c1}$ but below $P_{c}$. DW$_{0}$ stands for the uniform fully gapped DW.
DW$_{1}$ denotes the DW state when the imperfect nesting term $%
2t_{b}^{\prime }>\Delta _{0}$, so that the ungapped FS pockets or nonuniform
DW structure appear. }
\label{FigPhDia}
\end{figure}

The quasi-1D free electron dispersion without magnetic field has the form
\begin{equation}
\varepsilon (\boldsymbol{k})=\pm v_{F}(k_{x}\mp k_{F})+t_{\perp }(\mathbf{k}%
_{\perp }),  \label{1}
\end{equation}%
where the interchain dispersion $t_{\perp }({\boldsymbol{k}}_{\perp })$ is
much weaker than the in-plane Fermi energy $E_{F}\sim v_{F}k_{F}$ and given
by the tight-binding model with few leading terms:
\begin{equation}
t_{\perp }({\boldsymbol{k}}_{\perp })=2t_{b}\cos (k_{y}b)+2t_{b}^{\prime
}\cos (2k_{y}b).  \label{dispersion}
\end{equation}%
Here $b$ is the lattice constants in the $y$-direction, and usually $%
t_{b}^{\prime }\ll t_{b}\ll t_{a}\sim v_{F}k_{F}$. The dispersion along the $%
z$-axis is considerably weaker than along the $y$-direction and is omitted.
The FS consists of two warped sheets and possesses an approximate nesting
property, $2\varepsilon _{+}(\boldsymbol{k})\equiv \varepsilon (\boldsymbol{k%
})+\varepsilon (\boldsymbol{k}\mathbf{-}\boldsymbol{Q}_{N})\sim
t_{b}^{\prime }\ll E_{F}$, with $\boldsymbol{Q}_{N}\approx \left( 2k_{F},\pi
/b\right) $\ being the nesting vector. The nesting property leads to the
formation of DW at low temperature and is only spoiled by the second term $%
t_{b}^{\prime }(k_{y}\mathbf{)}$ in Eq. (\ref{dispersion}), which,
therefore, is called the "anti-nesting" term. Increase of the
latter with applied pressure leads to the transition in the
DW$_{1}$ state at $P>P_{c1}$ (see Fig. \ref{FigPhDia} below or
Fig. 1 in Refs. \cite{GrigorievPRB2008,GrigPhysicaB2009}), where
the quasi-particle states on the Fermi level first appear and lead
to the metallic conductivity or to SC
instability at low temperature $T<T_{c}^{SC}$. In the pressure interval $%
P_{c1}<P<P_{c}$ the new state develops, where the DW coexists with
superconductivity at rather low temperature $T<T_{c}^{SC}$, while
at higher temperature $T_{c}^{SC}<T<T_{c}^{DW}$ the DW state
coexists with the metallic phase. This coexistence takes place via
macroscopic phase separation \cite{ChaikinPRL2014}, via the
formation of small ungapped pockets \cite{GrigorievPRB2008} or via
the soliton phase \cite{GL,GG}. In most DW superconductors, the DW
transition temperature is much greater than the SC
transition temperature, $T_{c}^{DW}\gg T_{c}^{SC}$. For example, in (TMTSF)$%
_{2}$PF$_{6}$ $T_{c}^{SDW}\approx 8.5K\gg T_{c}^{SC}\approx 1.1K$, and in $%
\alpha $-(BEDT-TTF)$_{2}$KHg(SCN)$_{4}$, $T_{c}^{CDW}\approx 8K\gg
T_{c}^{SC}\approx 0.1K$. Below we consider the temperature interval $%
T_{c}^{SC}<T<T_{c}^{DW}$, i.e. the state with metallic conductivity.

In the $\tau $ -approximation, the conductivity along the main axes is given
by\cite{Abrik}:
\begin{equation}
\sigma _{i}\left( T\right) =e^{2}\tau \sum_{\boldsymbol{k}}\,v_{i}^{2}\left(
\boldsymbol{k}\right) \left( -n_{F}^{\prime }\left[ \varepsilon \left(
\boldsymbol{k}\right) \right] \right) ,  \label{sT}
\end{equation}%
where $e$ is the electron charge, $\tau $ is the mean free time, $%
\boldsymbol{k}$ is electron momentum, $v_{i}$ is the component of the
electron velocity along the $i$ -direction, $n_{F}^{\prime }(\varepsilon
)=-1/\{4T\cosh ^{2}\left[ (\varepsilon -\mu )/2T\right] \}$ is the
derivative of the Fermi distribution function, which restricts the summation
over momentum to the vicinity of FS, $\mu $ is the chemical potential, and $%
\varepsilon \left( \boldsymbol{k}\right) $ is the electron
dispersion. The electron velocity is calculated from Eq. (\ref{1})
using
\begin{equation}
v_{i}\left( \boldsymbol{k}\right) =\partial \varepsilon \left( \boldsymbol{k}%
\right) /\partial k_{i}.  \label{v}
\end{equation}%
Since the electron dispersion differs for various DW structures, so does the
temperature dependence of conductivity anisotropy $\sigma _{i}/\sigma _{j}$.
Now we consider this anisotropy for the initial electron dispersion given by
Eqs. (\ref{1}) and (\ref{dispersion}).

\subsection{DW state with open FS pockets}

\begin{figure}[tb]
\includegraphics[width=0.49\textwidth]{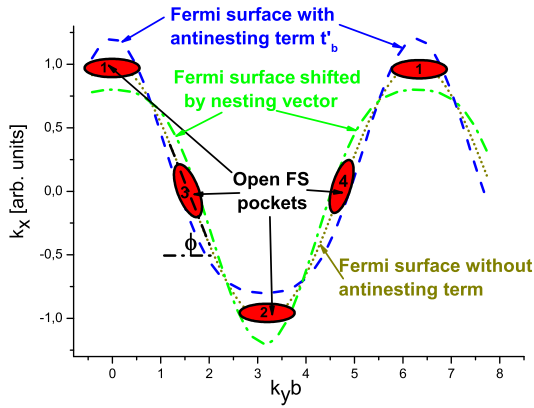}
\caption{(Color online) The schematic picture of small open pockets on one
Fermi surface sheet, which get formed when the anti-nesting term $%
2t_{b}^{\prime }$ in Eq. (\protect\ref{dispersion})\ exceeds the DW energy
gap $\Delta _{0}$. The blue dashed line shows the Fermi surface sheet with
imperfect nesting, i.e. with $2t_{b}^{\prime }>\Delta _{0}$. The green
dash-dotted line shows the other Fermi surface sheet shifted by the nesting
vector. If the nesting was perfect, these two lines would coincide. The
dotted brown line shows the perfectly nested Fermi surface sheet. The red
solid ellipses are the small Fermi surface pockets, that appear in the DW
state when $2t_{b}^{\prime }>\Delta _{0}$, i.e. when pressure exceeds $%
P_{c1} $.} \label{FigPockets}
\end{figure}

The electron dispersion in the DW state with imperfect nesting and open FS
pockets for the initial electron dispersion given by Eqs. (\ref{1}) and (\ref%
{dispersion}) was studied in Ref. \cite{GrigorievPRB2008}. There are eight
elongated open FS pockets in this case, four pockets per each FS sheet (see
Fig. \ref{FigPockets} or Fig. 2 in Ref. \cite{GrigorievPRB2008}). The
in-plane conductivity anisotropy is determined by the two inclined FS
pockets, marked by numbers 3 and 4, in this figure. The tangent of the
inclination angle $\phi $ of these pockets (see Fig. \ref{FigPockets})
approximately gives the ratio $\left\langle v_{y}^{2}\right\rangle
/\left\langle v_{x}^{2}\right\rangle \sim \tan ^{2}\phi \approx \left(
t_{b}/t_{a}\right) ^{2}$, where the angular brackets mean averaging over FS.
Thus, in the DW state one expects the anisotropy ratio
\begin{equation}
\sigma _{y}/\sigma _{x}\approx \left\langle v_{y}^{2}\right\rangle
/\left\langle v_{x}^{2}\right\rangle \sim \tan ^{2}\phi \sim \left(
t_{b}/t_{a}\right) ^{2} \ll 1,  \label{sa1}
\end{equation}%
which is close to the anisotropy without DW and slightly depends on
pressure. The change of this anisotropy ratio due to the transition to DW is
small for this scenario of the microscopic DW structure.

\subsection{Soliton-wall density-wave state}

In the soliton phase the DW order parameter depends on the
coordinate along the conducting chains: $\Delta \left( x\right)
\approx \Delta _{0}sn\left( x/\xi _{DW}\right) $, where
${\text{sn}}(y)$ is the elliptic sinus function. As a result an
array of soliton walls of width $\xi _{DW}=\hbar v_{F}/\pi \Delta
_{0}$ gets formed, where the DW order parameter changes sign. Each
soliton wall contributes one electron-like quasiparticle per
conducting chain on the Fermi level. For rather high soliton-wall
linear concentration $n_{s}$ the quasiparticles on different
soliton walls couple and form new conducting band in the middle of
the DW energy gap. The dispersion in this soliton band is
\cite{BGL,GG}
\begin{equation}
E\left( \mathbf{k}\right) =E\left( k_{x}\right) +\varepsilon _{+}(k_{y}),
\label{SolDisp}
\end{equation}%
where the interchain part of the dispersion is given by the anti-nesting
term in the dispersion (\ref{dispersion}):
\begin{equation}
\varepsilon _{+}(\mathbf{k}_{\perp })=\left[ t_{\perp }({\mathbf{k}}_{\perp
})+t_{\perp }({\mathbf{k}_{\perp }-\mathbf{Q}_{\perp }})\right] /2\approx
2t_{b}^{\prime }\cos (2k_{y}b).  \label{ep}
\end{equation}%
\qquad\ The dispersion $E\left( k_{x}\right) $ along the conducting chains
was found in Ref. \cite{BGK} (see Fig. 1 in Ref. \cite{BGK}), and for
qualitative analysis it can be approximated by%
\begin{equation}
E\left( k_{x}\right) \approx E_{-}\sin \left[ \pi \left( \left\vert
k_{x}\right\vert -k_{F}\right) /2\kappa _{0}\right] .  \label{EsAppr}
\end{equation}%
The soliton band width $E_{-}$ and boundary $\kappa _{0}=\pi n_{s}/2$ in the
momentum space are related to the linear concentration $n_{s}$ of the
soliton walls\cite{BGK}, which strongly depends on the anti-nesting electron
dispersion and, hence, on applied pressure. At small soliton concentration $%
n_{s}\rightarrow 0$ (see Ref. \cite{BrazKirovaReview}, p. 165):
\begin{equation}
E_{-}\approx 4\Delta _{0}\exp \left( -\Delta _{0}/\hbar v_{F}n_{s}\right) ,
\label{Em}
\end{equation}%
while at large soliton concentration $n_{s}\sim 1/\xi _{DW}$ $E_{-}\sim
\Delta _{0}$. To find $E_{-}$ and its pressure dependence one need to
minimize the total energy of soliton phase (see Appendix).

Substituting Eqs. (\ref{SolDisp})-(\ref{EsAppr}) to Eqs. (\ref{v}) and (\ref%
{sa1}) we obtain the mean square velocity in the soliton phase%
\begin{equation}
\left\langle v_{y}^{2}\right\rangle \approx 2\left( t_{b}^{\prime }b/\hbar
\right) ^{2},~\left\langle v_{x}^{2}\right\rangle \approx \left( \pi
E_{-}/2\hbar \kappa _{0}\right) ^{2}/2,  \label{vs2}
\end{equation}%
and the conductivity anisotropy%
\begin{equation}
\sigma _{y}/\sigma _{x}\approx \left\langle v_{y}^{2}\right\rangle
/\left\langle v_{x}^{2}\right\rangle \sim 16\left( t_{b}^{\prime }b\kappa
_{0}/\pi E_{-}\right) ^{2}.  \label{ans}
\end{equation}%
At $t_{b}^{\prime }\ll \Delta _{0}$ all electron states on the
Fermi level are gapped by a uniform DW, while at $t_{b}^{\prime
}\gg \Delta _{0}$ the normal-metal or electron-pocket DW state
are, usually, more favorable than the soliton DW state. Hence,
typically, in the soliton-wall non-uniform DW state $t_{b}^{\prime
}\sim \Delta _{0}$. Then the ratio $\sigma _{y}/\sigma _{x}\sim
16\left( t_{b}^{\prime }b\kappa _{0}/\pi E_{-}\right) ^{2}\gtrsim
1$ even at large soliton concentration $n_{s}$, when $E_{-}\sim
\Delta _{0}\sim 2t_{b}^{\prime }$. Hence, contrary to the
FS-pocket scenario, the formation of DW with soliton structure
leads to the strong change and even to the inversion of the
in-plane conductivity anisotropy. This allows to distinguish
experimentally these two microscopic structures of the DW with
imperfect nesting.

\subsection{Macroscopic phase separation}

For metal-DW phase separation in the form of macroscopic domains,\cite%
{ChaikinPRL2014} the calculation of conductivity anisotropy depends strongly
on the shape, size and mutual orientation of these domains. If the metallic
domains are weakly connected, one encounters the percolation regime.
Therefore, it is impossible to propose a general formulas for conductivity
anisotropy in this regime without specification of the particular domain
geometry. However, the change of the domain shape from filaments to the
walls, proposed in Refs. \cite{KangPRB2010,ChaikinPRL2014} to explain the $%
T_{c}^{SC}$ anisotropy, also affects the metallic conductivity anisotropy,
which can be easily measured. This issue requires additional experimental
and theoretical investigation.

\section{Conclusions}

We have shown that the change of conductivity anisotropy during
the transition to the DW state can reveal the microscopic
structure of this DW state. This is important in the case of
imperfect nesting, when the DW state has metallic conductivity due
to the ungapped electronic states on the Fermi level. In
particular, for the soliton DW
structure\cite{BrazKirovaReview,SuReview,BGL,GL,GG,GGPRB2007} the
DW transition leads to the strong change and sometimes to the
inversion of the in-plane conductivity anisotropy [see Eq.
(\ref{ans})]. On contrary, for the FS-pocket
scenario\cite{GrigorievPRB2008} of metal-DW coexistence the
in-plane anisotropy only slightly changes. This allows to
experimentally distinguish these two microscopic DW structures,
which may be useful to
describe the electronic properties of organic metals, such as $\alpha $-(ET)$%
_{2}$MHg(SCN)$_{4}$ with M = K or Tl and (TMTSF)$_{2}X$ with $X=$PF$_{6}$ or
ClO$_{4}$, and of various high-temperature superconductors, where
charge/spin-density wave coexists with superconducting/metallic state. For
macroscopic phase separation the conductivity anisotropy strongly depends on
the geometry and concentration of the metallic domains, which requires
investigation of the particular compounds.

\medskip

The work was supported by SIMTECH Program (Grant No. 246937), by
RFBR Grant No. 13-02-00178, by the Theoretical Center for the
Physics in Grenoble, and by the program '' Strongly correlated
electron-phonon systems'' of the Physics Branch of RAS.

\appendix

\section{Energy of soliton phase and the pressure dependence of the
electron-band width in this phase}

The energy of the soliton phase is (see Eqs. (33)-(35) of Ref.
\cite{BGL} or Eqs. (9)-(12) of Ref. \cite{GG})
\begin{equation}
W_{SP}=-\frac{\Delta _{0}^{2}}{2\pi \hbar v_{F}}+n_{s}A(t_{b}^{^{\prime
}})+n_{s}E_{-}^{2}B,  \label{13}
\end{equation}%
where the energy cost of a soliton wall per chain is (we denote $\varepsilon
_{\pm }(\mathbf{k}_{\perp })=\left[ t_{\perp }({\mathbf{k}}_{\perp })\pm
t_{\perp }({\mathbf{k}_{\perp }-\mathbf{Q}_{\perp }})\right] /2$):
\begin{equation}
A(t_{b}^{\prime })=(2/\pi )\Delta _{0}-2\int_{t_{\perp }\leq 0}\varepsilon
_{+}(k_{y})bdk_{y}/2\pi ,  \label{12}
\end{equation}%
and the interaction between the soliton walls is given by
\begin{equation}
B(t_{b}^{\prime })\approx \frac{1}{2\pi \Delta _{0}}-\sum_{FS}\frac{b/2\pi }{%
|d\varepsilon _{+}/dk_{y}|_{0}},  \label{15}
\end{equation}%
where $|d\varepsilon _{+}/dk_{y}|_{0}$ is the value of the transverse
velocity at the four values of $p_{\perp }$ where $t_{b}^{^{\prime
}}(k_{y})=0$. At $B>0$, crossing the point $A(t_{b}^{^{\prime }})=0$
corresponds to the second-order transition from the \textquotedblright
homogeneous\textquotedblright\ SDW to the soliton walls lattice with the
soliton wall concentration $n_{s}$ gradually increasing from zero. Negative $%
B<0$ would mean an abrupt first-order phase transition at $P=P_{c1}$, when
the soliton concentration $n_{s}$ jumps to some finite value.\cite{BGL} The
soliton band boundary in the momentum space has linear smallness, $\kappa
_{0}=\pi n_{s}/2,$ and the soliton band width is exponentially small for
small $n_{s}$ Differentiating of Eq. (\ref{13}) with respect to $E_{-}$
gives the optimal value of $E_{-}$ that minimizes the energy (\ref{13}):%
\begin{equation}
E_{-}^{2}=-A\left( t_{b}^{^{\prime }}\right) /\left[ B\left( t_{b}^{^{\prime
}}\right) \left( 2\ln \left( 4\Delta _{0}/E_{-}\right) +1\right) \right] .
\label{Emt}
\end{equation}%
This width depends strongly on the dispersion $\varepsilon _{+}(k_{y})$ in
the soliton band. For the dispersion (\ref{dispersion}) Eqs. (\ref{12}),(\ref%
{15}) give%
\begin{eqnarray}
A(t_{b}^{\prime }) &=&(2/\pi )\left( \Delta _{0}-2t_{b}^{\prime }\right) ,
\label{A} \\
B(t_{b}^{\prime }) &=&\frac{1}{2\pi \Delta _{0}}-\frac{1}{4\pi t_{b}^{\prime
}}.  \label{B}
\end{eqnarray}%
The occasional degeneracy, meaning that both $A(t_{b}^{\prime })$ and $%
B(t_{b}^{\prime })$ become zero at the same point $2t_{b}^{\prime }=\Delta
_{0}$, is a consequence of the particular electron dispersion (\ref%
{dispersion}). In real compounds this degeneracy is always removed by higher
harmonics in the dispersion (\ref{dispersion}). In the presence of the
occasional degeneracy (\ref{B}), the width of the soliton band at $%
P\rightarrow P_{c1}$ (i.e., at $\Delta _{0}\rightarrow 2t_{b}^{\prime }$)
reduces very slowly:
\begin{equation}
E_{-}\approx \frac{2\Delta _{0}}{\sqrt{2\ln \left( 4\Delta _{0}/E_{-}\right)
+1}}\sim \Delta _{0},  \label{Emd}
\end{equation}
which means a sharp (though second-order) transition from the uniform DW to
the soliton phase. Without the occasional degeneracy (\ref{A}),(\ref{B})%
\begin{equation}
E_{-}\sim \sqrt{\Delta _{0}\delta }\equiv \sqrt{\Delta _{0}\left(
2t_{b}^{\prime }-\Delta _{0}\right) }\propto \sqrt{P-P_{c1}}.  \label{Emnd}
\end{equation}%
A quantitative estimate of the dependence $E_{-}\left( P\right) $ requires a
more accurate calculation of the interaction (\ref{15}) between soliton
walls.
\bibliographystyle{elsarticle-num}
\bibliography{martin}

\end{document}